\documentclass{elsart}

\usepackage{epsfig}

\usepackage{amsmath,amssymb,amsfonts}
\usepackage{latexsym,exscale}

\begin{document}

\begin{frontmatter}



\title{Elliptic flow of multi-strange particles: 
fragmentation, recombination and hydrodynamics}


\author[label1]{Chiho Nonaka},  
\ead{nonaka@phy.duke.edu}
\author[label1]{Rainer J.~Fries}, 
\author[label1,label2]{Steffen A.~Bass}

\address[label1]{Department of Physics, Duke University, Durham, NC 27708, USA}
\address[label2]{RIKEN BNL Research Center, Brookhaven National Laboratory, 
        Upton, NY 11973, USA}
\begin{abstract}
We study the elliptic flow $v_2$ of multi-strange hadrons such as the
$\phi$, $\Xi$ and $\Omega$ as a function of transverse momentum
in the recombination and fragmentation model and compare to a standard
hydrodynamic calculation. 
We find that the measurement of $v_2$ for the $\phi$ and $\Omega$
will allow for the  unambiguous distinction 
between parton recombination and statistical 
hadro-chemistry to be the dominant process in hadronization
at intermediate transverse momenta.
\end{abstract}

\begin{keyword}
Relativistic heavy-ion collisions \sep Elliptic flow 
\PACS 25.75.Dw \sep 
25.75.Ld \sep 
24.85.+p 
\end{keyword}

\end{frontmatter}

\section{Introduction}
Elliptic flow of hadrons -- more precisely 
defined as the second Fourier coefficient of their azimuthal momentum 
distribution --
has been suggested as a sensitive probe 
for the buildup of pressure in the early reaction stage of relativistic
heavy-ion collisions \cite{v2_original}.
Data from Au+Au collisions at the Relativistic Heavy Ion Collider (RHIC) 
has revealed unexpected features in the behavior of the elliptic 
flow $v_2$  as a function of transverse momentum $P_T$: 
for non-central collisions $v_2$ first rises and then saturates at higher 
$P_T$ \cite{PHENIXv2,Esumi,STAR-v2,So,STAR-v2b}, the saturation occurring
at higher $P_T$ and
at higher values for $v_2$ for baryons than for mesons.

It was first pointed out by Voloshin that this surprising behavior 
can be explained 
in a naive parton recombination model for hadronization \cite{Vo}. 
Several more detailed studies on
elliptic flow based on recombination have been carried out subsequently
\cite{FrMuNoBa1,FrMuNoBa2,LiKo,MoVo,LiMo,GrKoLe2}.      
In addition to elliptic flow, parton recombination has been very
successful in explaining the behavior of particle spectra and ratios
at intermediate transverse momenta.
The general theory of recombination as a hadronization mechanism in dense 
parton systems is described in great detail in
\cite{FrMuNoBa1,FrMuNoBa2,GrKoLe2,BiLeZi95,BiLeZi98,ZiBiCsLe,CsLe,HwYa,GrKoLe1}.
In this letter we present quantitative results for the elliptic flow of 
multi-strange particles in the recombination plus fragmentation approach
-- the theoretical foundation being based on 
our previous work in \cite{FrMuNoBa2}.

At low $P_T$ the experimental data show that the elliptic flow of a heavy 
particle is smaller than that of a light particle.   
This mass-dependence can be understood in hydrodynamic 
models where particles with higher mass 
experience the collective movement of the medium less than lighter species
\cite{KoHuHe,HuKoHe}.
However, the saturation of $v_2$ above $P_T \approx 1.5$ GeV/$c$ 
and its dependence on the
hadron species in this region contradicts a purely
hydrodynamic model.

At high $P_T$, the energy loss of fast partons in the dense medium has been 
proposed as a mechanism that can translate the anisotropy in the initial
Au+Au system into an momentum anisotropy in the final state as well 
\cite{Wa1,GyViWa,Wa2}. The observation of jet quenching at RHIC energies is 
a strong indication for the existence of partonic energy loss 
which may lead to such an azimuthal anisotropy \cite{PHENIX-su}.

In the most interesting region of intermediate $P_T$, between 2 and 5 GeV/$c$, 
the use of hydrodynamics becomes questionable, since the fundamental 
assumption of an
infinitely small mean free path is not valid anymore. 
On the other hand, the use of
perturbative QCD is not well justified either, since
the relevant scale $\sim P_T$ is
rather small and may lead to uncontrollable higher twist corrections. 
In addition,
the strong energy loss further suppresses the contribution from perturbative
parton fragmentation. These considerations suggest 
recombination as the dominant hadronization 
mechanism in the intermediate $P_T$ region.

To clarify the situation and to establish the dominant mechanism for 
hadronization and the
creation of hadron $v_2$ in the intermediate $P_T$ domain we propose the
measurement of $v_2$ for multi-strange hadrons, in particular for the
$\phi$ and the $\Omega$.
In a hydrodynamic picture the differences between hadron 
species are predominantly driven by the mass differences. In the recombination 
formalism the mass only plays a role at low $P_T$, while strong deviations 
appear at
intermediate $P_T$ due to the different valence structure of baryons and 
mesons. Since the mass of the $\phi$ 
is very close to that of the $\Lambda$, a statistical hadronization scheme
as employed by hydrodynamics would predict
the $\phi$ to behave similar to the $\Lambda$, while recombination
implies that it behaves similar to other mesons, e.g. the kaon.

\section{Recombination plus fragmentation approach}
In \cite{FrMuNoBa1,FrMuNoBa2} we found that
hadron production at intermediate $P_T$ is governed by an interplay
of fragmentation and recombination. 
Recombination is more efficient than fragmentation 
if the parton spectrum is exponential, while fragmentation dominates for 
the (perturbative) power law tail of the partonic $P_T$ spectrum.
E.g. for mesons $M$ recombining on a hypersurface $\Sigma$, the momentum 
distribution is given by
\begin{equation}
  E \frac{N_M}{d^3P}=C_M \int_\Sigma d \sigma_R 
  \frac{P \cdot u(R)}{(2 \pi)^3}\int^1_0dx \\ 
  w_a(R;x P^+) |\phi_M(x)|^2 w_b(R;(1-x)P^+), 
\label{Eq-rmeson}
\end{equation}
where $\phi_M(x)$ is a wave function for the meson on the light cone, $x$
is the momentum fraction of valence quark $a$, the $w$ are the
momentum distributions of the quarks $a$ and $b$ and $C_M$ is a degeneracy 
factor. We refer the reader to \cite{FrMuNoBa2} for more details.

The parton momentum distribution 
at hadronization is an input in Eq.\ (\ref{Eq-rmeson}).
For quantitative predictions we assume a thermal distribution,    
\begin{equation}
  w_a(R;p)=\gamma_ae^{-p\cdot v(R)/T}e^{-\eta^2/2 \Delta^2}f(\rho, \phi), 
\label{Eq-rpa}
\end{equation}
where $v(R)$ is a four velocity and $\gamma_a$ is fugacity factor for each 
parton species $a$. The longitudinal and transverse spatial distributions are 
described by the width $\Delta$ for the  space-time rapidity $\eta$ and a 
profile function $f(\rho, \phi)$ for transverse size and azimuthal angle
$\rho$ and $\phi$. We set $\Delta=2$. $f(\rho, \phi)$ is impact parameter
dependent and is scaled with the volume of the overlap zone of two nuclei 
in non-central collisions. 

We assume that the hadronization process takes place at a temperature 
$T = 175$ MeV on a hypersurface $\Sigma$ given by 
$\tau= \sqrt{t^2-z^2}=5$ fm/$c$. The radial flow velocity $v_T$ is fixed at
0.55$c$ for all values of $\rho$ and for all impact parameters. These input 
parameters were determined in \cite{FrMuNoBa2} 
and are consistent with
the observed hadron spectra and ratios from PHENIX and STAR.
The recombination calculation has to be supplemented by a pQCD calculation
using fragmentation functions. However, no information is available concerning
the  fragmentation functions of strange hadrons 
beyond kaons and $\Lambda$s.

\section{Elliptic flow}
The azimuthal asymmetry $v_2$ is defined by   
\begin{equation}
  v_2(P_T) = \langle \cos 2 \Phi \rangle 
  = \frac{\int d \Phi \cos 2 \Phi d^2 N/d^2P_T}
  {\int d \Phi d^2N/d^2P_T},
\label{Eq-defv2}
\end{equation}
where $\Phi$ is the azimuthal angle in momentum space. 

At low and intermediate $P_T$ the initial geometric anisotropy results in an 
anisotropic pressure gradient that implies an elliptic flow profile. This is 
para\-metrized  via an anisotropic parton momentum distribution 
at the time of hadronization at which partons then 
hadronize via recombination. We use Eq.\ (\ref{Eq-rpa}) together with
an elliptic modulation of the transverse rapidity
\begin{equation}
  \eta_T (\phi,p_T) = \eta^0_T(1-F(p_T)\cos 2 \phi)
  \label{eq:p1}
\end{equation}
to determine $v_2$ for the partons at hadronization.  
$\eta^0_T$ is the rapidity given by the radial flow velocity $v_T=0.55 c$,
so that $\tanh \eta_T^0 = 0.55$.
This is inspired by a hydrodynamical description of the parton phase at low
$p_T$.
Hydrodynamics works well to describe the measured $v_2$ for hadrons up 
to $P_T=1.5$ GeV/$c$ \cite{STAR-v2,TeLaSh,Hi,HeKo}. 
In order to accommodate the deviation from ideal hydrodynamics at higher 
$P_T$ we assume that $F(p_T)$ takes the from
\begin{equation}
  F(p_T) = \frac{\alpha}{1+(p_T/p_0)^2},
  \label{eq:p2}
\end{equation}
where for a given impact parameter $b$
\begin{equation}
 \alpha = \frac{w(b)-l(b)}{w(b)+l(b)}
\end{equation}
is determined by the collision geometry with width
$w(b)$=$R_A-b$ and length $l(b)=\sqrt{R^2_A-(b/2)^2}$ of the collision
zone. $R_A$ is the radius of a gold nucleus.

We assume that $v^u_2=v^{\bar{u}}_2=v^d_2=v^{\bar{d}}_2$ and 
$v^s_2=v^{\bar{s}}_2$. The slight difference between light quarks and strange 
quarks originates from the mass difference. We use constituent masses 
$m_{u,d}=260$ MeV and
$m_s= 460$ MeV. The parameter $p_0=1.1$ GeV/$c$ was obtained in 
\cite{FrMuNoBa2} by a fit to the elliptic flow of pions obtained by PHENIX
\cite{PHENIXv2} and is also consistent with data on $v_2$ for charged hadrons,
protons, kaons and $\Lambda$s \cite{FrMuNoBa2}.

In \cite{FrMuNoBa2} we showed that for recombining mesons $M$ and 
baryons $B$ the anisotropy $v_2$ can be written as
\begin{equation}
  v^M_2(P_T) = \frac{\int dx |\phi_M(x)|^2  
  [ v^a_2(x P_T) + v^b_2((1-x)P_T)]k_M(x,P_T)}
  {\int dx |\phi_M(x)|^2[1+2v^a_2(xP_T)v^b_2((1-x)P_T)]k_M(x,P_T)}, 
  \label{Eq-mesonv2}
\end{equation}
\begin{small}
\begin{equation}
  v^B_2(P_T) = 
 \frac{\int {\mathcal D}x_i  
   |\phi_B(x_i)|^2[ \sum_{j=a,b,c} v^j_2(x_j P_T) + 
   3\prod_{j=a,b,c} v^j_2(x_j P_T)] k_B(x_i,P_T)}
   {\int {\mathcal D}x_i | \phi_B(x_i)|^2
   [1 + 2\sum_{j=a,b,c}  \prod_{\mu\not=j} v^\mu_2(x_\mu P_T)
   ] k_B(x_i,P_T)}, 
  \label{Eq-baryonv2}
\end{equation}
\end{small}
where $k_M(x, P_T)$ and $k_B(x,P_T)$ are given by 
\begin{equation}
  k_M(x,P_T)=K_1\left [
  \frac{\cosh \eta_T}{T} \left (
  \sqrt{m^2_a+x^2P^2_T} + \sqrt{m^2_b+(1-x)^2P^2_T}
  \right )  
  \right ], 
\end{equation}
\begin{equation}
  k_B(x_i,P_T)=K_1\left [
  \frac{\cosh \eta_T}{T} \sum_{j=a,b,c}
  \sqrt{m^2_a+x^2_a P^2_T} 
  \right ].
\end{equation}
$a$, $b$ and $c$ stand for the valence partons respectively.
We use the short notation $\int {\mathcal D}x_i=\int^1_0 dx_adx_bdx_c 
\delta(x_a+x_b+x_c-1)$ in Eq.\ (\ref{Eq-baryonv2}).
Using infinitely narrow delta shaped wave functions $|\phi|^2\sim \delta$ that
distribute the momentum of the hadron equally to all valence quarks and
neglecting higher order terms one obtains the simple scaling law 
\cite{So,STAR-v2b,Vo,FrMuNoBa2,MoVo,LiMo}, 
\begin{equation} 
  v^h_2(P_T) = nv^a_2 \left ( \frac{1}{n}P_T \right ), 
  \label{eq:scaling}
\end{equation}
where $n$ is the number of valence quarks in the hadron.

At high $P_T$ the elliptic flow is dominated by the contribution of 
fragmentation and its value is determined by the azimuthal anisotropy due to
jet energy loss. Since fragmentation functions of multi-strange particles are
basically unknown, a precise calculation is not possible. However, the energy
loss and hence the resulting anisotropy are partonic in nature and one can
argue that in the ratio of Eq.\ (\ref{Eq-defv2}) the effects of fragmentation 
functions tend to cancel. Indeed, we have shown in \cite{FrMuNoBa2} that
for the part of $v_2$ coming from fragmentation the difference between 
different hadron species is numerically negligible. Keeping this in mind
we approximate the component of $v_2$ coming from fragmentation for baryons
and mesons by that of $\Lambda$s and kaons respectively.

The total elliptic flow can be written as   
\begin{equation}
  v_2(P_T)=r(P_T)v_{2,R}(P_T) + (1 - r(P_T))v_{2,F}(P_T),
\end{equation}
where $r(P_T)$ is the ratio recombination / (recombination + fragmentation)
for the hadron spectrum as a function of transverse momentum.

\section{Numerical results}
\begin{figure}
\includegraphics*[width=0.9\columnwidth]{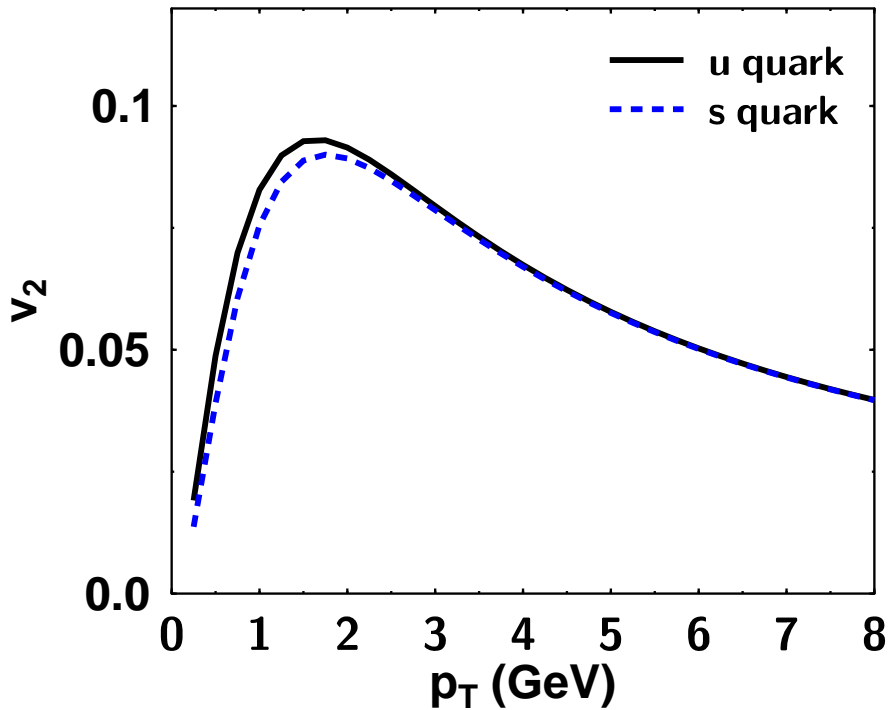}
  \caption{\label{Fig-v2pa} Elliptic flow of $u$ quarks and $s$ quarks as a 
  function of $p_T$ (thermal phase only). }
\end{figure}

In the following we shall discuss our results for the 
elliptic flow of (multi-)strange particles.
All results shown are for Au+Au collisions at
$\sqrt{s}= 200$~GeV with an impact parameter $b$ of 8.0 fm.
Starting point of our calculation is a system of 
constituent quarks prior to hadronization. 
Figure \ref{Fig-v2pa} shows $ v_2$ of $u$ and $s$ quarks as a 
function of $p_T$ from Eqs.\ (\ref{Eq-defv2},\ref{eq:p1},\ref{eq:p2}). 
The slight difference between $u$ and $s$ quarks stems from their 
different masses. The mass effect disappears at high $p_T$.   
The assumption of equal elliptic flow
of $s$ and $u,d$ quarks is well justified given the nearly identical
behavior of the kaon and pion elliptic flow as a function of $p_T$,
as can be seen in \cite{FrMuNoBa2}. Note that a different value for the 
$s$ quark elliptic flow will lead
to a characteristic flavor ordering for the elliptic flow among multi-strange
mesons and baryons respectively, depending on their strange quark content 
\cite{LiKo,MoVo,LiMo}. Such an effect has not been observed in the data so far.

\begin{figure}
\includegraphics[width=0.9\columnwidth]{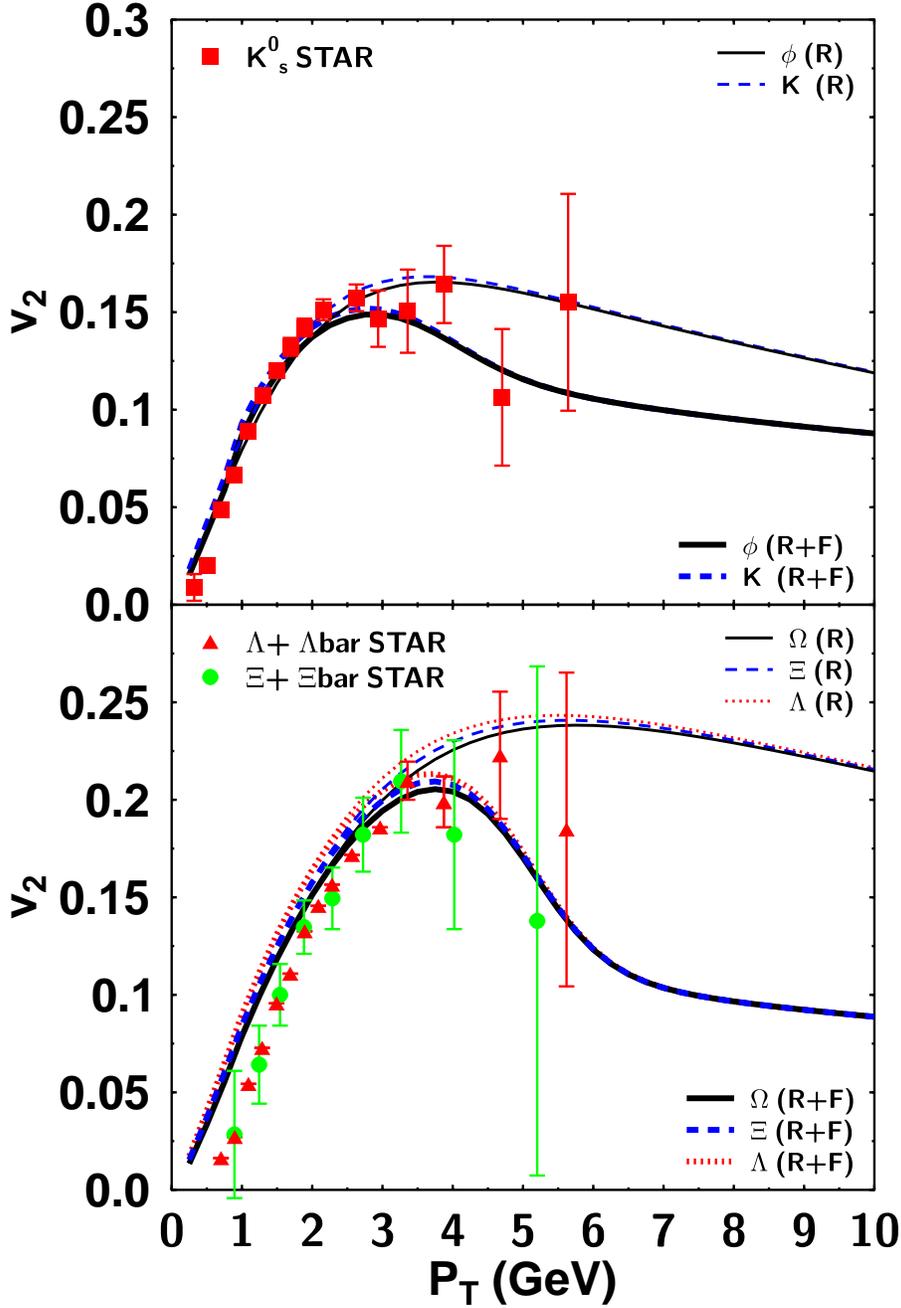}
  \caption{\label{Fig-v2ms}
  Elliptic flow for multi-strange mesons (top frame) 
  and baryons (bottom frame) in the recombination+fragmentation
  approach compared to data from STAR (symbols) ($\Lambda+\bar{\Lambda}$ 
\cite{STAR-v2b}, $\Xi +\bar{\Xi}$ \cite{Ca}). Thick lines denote the full
  calculation, whereas thin lines refer to the recombination contribution. }
\end{figure}

Figure \ref{Fig-v2ms} shows $v_2$ for multi-strange mesons (top frame)
and baryons (bottom frame) 
 as a function of $P_T$.  
The experimental data are 
taken from the STAR collaboration \cite{STAR-v2b,Ca}.
Thick lines denote the full recombination+fragmentation calculation (labeled
as R+F in the figure),
whereas thin lines refer to recombination from thermal quarks 
alone (labeled as R in the figure).
Under the assumption that the $s$ quark elliptic flow is identical to $u$, $d$
quark elliptic flow all mesons show the same behavior.
The $v_2$ for $\phi$ is identical to that of the $K$. 
This is most prominent in the saturation region of $v_2$. In the 
recombination approach the value of $v_2$ there is solely determined by the 
number of valence quarks of the hadron and the partonic $v_2$.
The mass effect which drives the $v_2$ characteristics in 
hydrodynamical calculations is negligible for the saturation feature. 
The same systematics can be observed for baryons as well, leading
to nearly identical elliptic flow values for $\Lambda$, $\Xi$ and $\Omega$. 
Experimental data confirm our calculation in the R+F approach 
for kaons, $\Lambda$s and $\Xi$ between 1.5 and 6 GeV/$c$. 
We have shown in \cite{FrMuNoBa2} that the dependence of elliptic flow on 
the shape of the wave function used in 
Eqs.\ (\ref{Eq-mesonv2},\ref{Eq-baryonv2})
is negligible for transverse momenta larger
than 2 GeV. For further studies on the influence of the wave function see 
e.g.\ \cite{BiLeZi98}.

At $P_T$ lower than $1.5$~GeV the data exhibit the well known splitting 
of the $v_2$ curves due to the mass of the hadrons. 
This is only poorly reproduced by the recombination formalism since 
binding energies are not properly taken into account, whereas hydrodynamic 
calculations perform very well in this domain. 
  
\begin{figure}
\includegraphics[width=0.9\columnwidth]{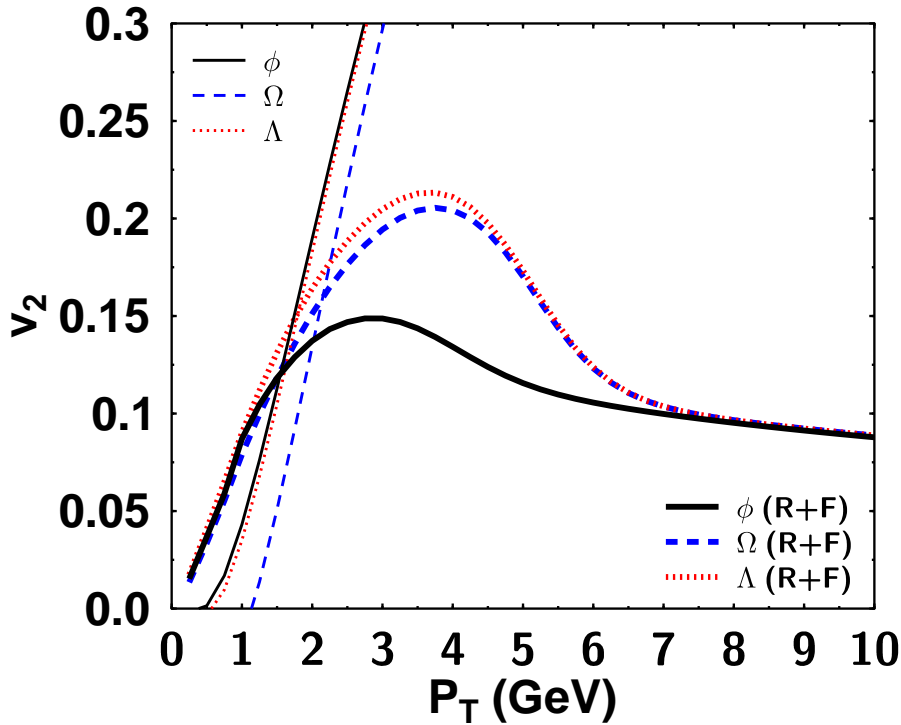}
  \caption{\label{hydro_reco}
  Elliptic flow for some strange hadrons 
  in the recombination+fragmentation approach (thick lines) and
  in a hydrodynamic calculation (thin lines).}
\end{figure}

The fragmentation contribution to $v_2$ and the ratio
$r(P_T)$ for kaons and $\Lambda$s are taken from \cite{FrMuNoBa2}. 
As discussed above we use
the values for kaons to estimate both $v_2$ from fragmentation and $r(P_T)$ 
for the $\phi$. Moreover we use the values for the $\Lambda$ 
to estimate the corresponding quantities for the $\Xi$ and the $\Omega$ 
in the fragmentation process. 
The tendency of fragmentation functions to cancel in Eq.\ (\ref{Eq-defv2})
makes this a good approximation and also makes the difference between mesons
and baryons vanish. This implies a universal flat behavior for 
$P_T> 6$ GeV/$c$ where recombination is suppressed and the azimuthal 
anisotropy is dominated by pQCD processes and parton fragmentation.
The merging of meson and baryon $v_2$ into one universal
curve at high $P_T$ can be seen even better in Fig.\ \ref{hydro_reco}.
%
%
Multistrange hadrons with high masses might be disfavored by the 
fragmentation process. We expect the relative weight of recombination
to be even larger for multi-strange particles: $r_\phi(P_T) > r_K(P_T)$,
$r_\Omega(P_T) > r_\Lambda(P_T) > r_\Xi(P_T)$. Therefore hadrons
like the $\Xi$ and the $\Omega$ might reach the universal curve for $v_2$ from 
fragmentation even later than indicated in Fig.\ \ref{Fig-v2ms}

Figure~\ref{hydro_reco} elucidates the systematic differences between $v_2$
originating from statistical hadronization in 
a hydrodynamic scenario (blast wave model 
\cite{HuKoHe,SiRa}) vs. hadron $v_2$ generated from a 
recombination+fragmentation scenario: thin lines represent a hydrodynamic
calculation of the elliptic flow of $\phi$, $\Lambda$ and $\Omega$ as a 
function of $P_T$. Since the masses of the $\phi$ and the $\Lambda$ are 
similar, their $v_2$ is identical as well in the 
hydrodynamic calculation. Due to its much larger mass the $\Omega$ exhibits
a smaller $v_2$, clearly distinct from that of the other two hadron species. 
The situation reverses in the region where recombination
dominates: there the number of valence quarks in
the hadron determines the resulting $v_2$, leading to a distinct difference
in the $v_2$ between the $\phi$ and the $\Lambda$, which now coincides with
the $\Omega$. At highest $P_T$ all curves merge to the universal $v_2$ from
fragmentation.

We assume a sudden freeze-out and neglect interactions in the hadronic phase
in both descriptions, recombination plus fragmentation and hydrodynamics.
This seems to be reasonable for particles with high and intermediate transverse
momentum and is supported by the success of the scaling law Eq.\ 
(\ref{eq:scaling}). 
The study of resonance particles is particularly important in this case
\cite{CsLe,BaDu}. We will investigate this in a forthcoming publication.

\section{Conclusion}
In brief, we investigated the elliptic flow of multi-strange hadrons 
in the recombination plus fragmentation approach. 
We find that the behavior of $v_2$ as a function of $P_T$ 
in this approach is dominated by the number of valence quarks of the
respective hadron, leading to very similar values of the elliptic flow
for all mesons and likewise for all baryons, 
nearly independent of the hadron mass, as opposed to conventional
statistical hadronization in hydrodynamic calculations.
In particular we find that $v_2$ for $\phi$ mesons is nearly identical 
to that of kaons above 2 GeV/$c$. This uses a parton $v_2$ for strange 
quarks that is for all practical purposes equal to the $v_2$ of the light
flavors, in accordance with RHIC data.
The measurement of $\phi$ and $\Omega$ elliptic flow will thus permit to 
distinguish between statistical hadronization as employed by standard 
hydrodynamic
calculations vs. hadron production through parton recombination at
intermediate $P_T$.
At high $P_T$ we find the elliptic flow to be dominated 
by fragmentation, leading to a 
universal curve above 6 GeV/$c$.

\section*{Acknowledgments}
This work was supported in part by RIKEN, the Brookhaven National 
Laboratory, and DOE grants DE-FG02-96ER40945 and DE-AC02-98CH10886.
S.A.B. acknowledges support from an Outstanding Junior Investigator
Award (DOE grant DE-FG02-03ER41239) and R.J.F. has been supported
by a Feodor Lynen Fellowship of the Alexander von Humboldt Foundation.




\end{document}